\documentclass{ws-rv975x65}
\usepackage{ws-rv-van}             
\makeindex
\usepackage{graphicx}
\usepackage{color}
\usepackage{bm}

\begin{document}

\chapter{Self-Consistent Theory of Anderson Localization:\\
General Formalism and Applications
\label{ch1}}

\author[P.~W\"olfle and D. Vollhardt]{P.~W\"olfle$^1$  and D. Vollhardt$^2$
}

\address{$^1$ Institute for Condensed Matter Theory,
Institute for Nanotechnology and DFG-Center for Functional Nanostructures,
Karlsruhe Institute of Technology, D-76131 Karlsruhe, Germany\\
$^2$Theoretical Physics III, Center for Electronic Correlations and Magnetism,
Institute for Physics, University of Augsburg, D-86135 Augsburg, Germany\\
}

\begin{abstract}
The self-consistent theory of Anderson localization of quantum particles or
classical waves in disordered media is reviewed. After presenting the basic
concepts of the theory of Anderson localization in the case of electrons
in disordered solids, the regimes of weak and strong localization are
discussed. Then the scaling theory of the Anderson localization transition
is reviewed. The renormalization group theory is introduced and results and
consequences are presented. It is shown how scale-dependent terms in the
renormalized perturbation theory of the inverse diffusion coefficient lead
in a natural way to a self-consistent equation for the diffusion
coefficient. The latter accounts quantitatively for the static and dynamic
transport properties except for a region near the critical point. Several
recent applications and extensions of the self-consistent theory, in particular for classical waves, are
discussed.
\end{abstract}

\body

\section{Introduction to Anderson localization}

\label{sec1.1}

\subsection{Brief historical review}

The localization of quantum particles by a static random potential, or of
classical waves by random fluctuations of the medium, is one of the most
intriguing phenomena in statistical physics. The key ingredient of
localization, wave interference, was introduced in P. W. Anderson's seminal
paper ``Absence of diffusion in certain random lattices''.~\cite{and158}
There it was shown that electrons may be localized by a random potential, so
that diffusion is suppressed, even in a situation where classical particles
would be delocalized. The fundamental reason for the localizing effect of a
random potential on quantum particles or classical waves is the multiple
interference of wave components scattered by randomly positioned scattering
centers. The interference effect takes place, as long as the propagation is
coherent.

It is interesting to note that the first application of the idea of
localization concerned the spin diffusion $D$ of electrons and not the
electrical conductivity $\sigma $. Anderson considered a tight-binding model
of electrons on a crystal lattice, with energy levels at each site chosen
from a random distribution.~\cite{and158} The traditional view had been,
that scattering by the random potential causes the Bloch waves to lose well-defined momentum
on the length scale of the mean-free path $\ell $. Nevertheless,
the wavefunction was thought to remain extended throughout the sample.
Anderson pointed out that if the disorder is sufficiently strong, the
particles may become localized, in that the envelope of the wave function $\psi (\bm{r})$ decays exponentially from some point $\bm{r}_{0}$ in space:
\begin{equation}
\mid \psi (\bm{r})\mid \sim \exp (\mid \bm{r}-\bm{r}_{0}\mid /\xi),
\label{1.1}
\end{equation} where $\xi $ is the localization length.

There exist a number of review articles on the Anderson localization
problem. The most complete account of the early work was presented by Lee
and Ramakrishnan~\cite{leer85}. The seminal early work on interaction
affects is presented in Ref.~\cite{alta85}. A complete account of the early
numerical work can be found in Ref.~\cite{kmso}. A path integral formulation
of weak localization is presented in Ref.~\cite{chaks86}. Several more
review articles and books are cited along the way. In the following we will
use units with Planck's constant $\hbar $ and Boltzmann's constant $k_{B}$\
equal to unity, unless stated otherwise.

\subsection{Electrons and classical waves in disordered systems}

The wavefunction $\psi (\bm{r})$ of a single electron of mass $m$ in a
random potential $V(\bm{r})$ obeys the stationary Schr\"{o}dinger equation
\begin{equation}
\Big(-\frac{\hbar ^{2}}{2m}\bm{\nabla}^{2}+V(\bm{r})-E\Big)\psi (\bm{r})=0.
\label{1.2}
\end{equation}%
In the simplest case $V(\bm{r})$ may be assumed to obey Gaussian statistics
with $\langle V(\bm{r})V(\bm{r^{\prime }})\rangle =\langle V^{2}\rangle
\delta (\bm{r}-\bm{r^{\prime }})$, but many of the results presented below
are valid for a much wider class of models. Electrons propagating in the
random potential $V(\bm{r})$ will be scattered on average after a time $\tau
$. For weak random potential the scattering rate is given by
\begin{equation}
\frac{1}{\tau }=\pi N(E)\;\langle V^{2}\rangle   \label{1.3}
\end{equation}%
where $N(E)$ is the density of states at the energy $E$ of the electron. In
a metal the electrons carrying the charge current are those at the Fermi
energy $E=E_{F}$. Within the time $\tau $ the electron travels a distance $\ell =v_{F}\tau $, where $v_{F}$ is its velocity.

In close analogy the wave amplitude $\psi (\bm{r})$ of a classical
monochromatic wave of frequency $\omega $ obeys the wave equation
\begin{equation}
\Big(\frac{\omega ^{2}}{c^{2}(\bm{r})}+\bm\nabla ^{2}\Big)\psi (\bm{r})=0.
\label{1.4}
\end{equation}%
Here $c(\bm{r})$ is the wave velocity at position $\bm{r}$ in an
inhomogeneous medium, assumed to be a randomly fluctuating quantity. The
main difference between the Schr\"{o}dinger equation and the wave equation
is that in the wave equation the \textquotedblleft random
potential\textquotedblright\ $1/c^{2}(\bm{r})$ is multiplied by $\omega ^{2}$, so that disorder is suppressed in the limit $\omega \rightarrow 0$. By
contrast, in the quantum case disorder will be dominant in the limit of low
energy $E$. A further difference may arise if the wave amplitude is a vector
quantity as, e.g., in the \ case of electromagnetic waves in $d=3$
dimensions.

In real systems particles or wave packets are not independent, but interact.
Electrons are coupled by the Coulomb interaction, leading to important
effects that go much beyond the single particle model. Similarly,
wave packets interact via nonlinear polarization of the medium.
Apart from these complications, the physics of electronic wave packets and
classical wave packets is quite similar. In the following we will present
most of the discussion in the language of electronic wave packets.

\subsection{Weak localization}

\label{Weak-localization}

The all-important effect of wave interference is most clearly seen in the
limit of weak scattering, where it already may cause localization, but only
in reduced dimensions. While it is difficult to observe full localization at
finite temperature $T$, on account of the effect of interactions limiting
the phase coherence, the dramatic signatures of localization are visible at
finite $T$ in the form of ``weak localization''.\cite{aalr79,gor79}

An electron or a wave packet moving through a disordered medium will be
scattered by the random potential on the average after propagating a
distance $\ell $, the mean-free path. On larger length scales the
propagation is diffusive. Weak localization is a consequence of destructive
interference of two wave components starting at some point and returning to
the same point after traversing time-reversed paths. Let the probability
amplitudes for the wave packet to move from point $\bm{r}_{0}$ along some
path $C_1$ back to $\bm{r}_{0}$ be $A_1$ and along a different path $C_2$ be $A_2$, then the transition probability for the particle to move either along $C_1$ or along $C_2$ will be
\begin{equation}
w=\mid A_{1}+A_{2}\mid ^{2}=w_{cl}+w_{int}\text{,}
\end{equation}%
where $w_{cl}=\mid A_{1}\mid ^{2}+\mid A_{2}\mid ^{2}$ and $w_{int}=2Re(A_{1}^{\ast
}A_{2})$. For any two paths the interference term $w_{int}$ may be positive
or negative, and thus averages to zero. However, if $A_2=A_r$ is the amplitude of the time-reverse of path
$A_1=A$ and if time reversal holds, then $A=A_r$, i.e., the probability of return $w$
is enhanced by a factor of two compared to the probability $w_{cl}$ of a
classical system:
\begin{equation}
w=4\mid A\mid ^{2}=2w_{cl}.  \label{1.6}
\end{equation}%
In that case the probability for transmission is reduced, which leads to a
reduced diffusion coefficient and a reduced conductivity. One may estimate
the correction to the conductivity in the following qualitative way. The
relative change of the conductivity $\sigma $ by the above interference
effect is equal to the probability of interference of two wave packets of
extension $\lambda $, the wavelength, after returning to the starting point.
The infinitesimal probability of return to the origin in time $t$ of a
particle diffusing in $d$ dimension is given by $(4\pi Dt)^{-d/2}d^{3}r$
where $D$ is the diffusion coefficient. Since the volume of interference in
the time interval $[t,t+dt]$ is $\lambda ^{d-1}vdt$, where $v$ is the
velocity of the wave packet, one finds the quantum correction to the
conductivity $\delta \sigma $ as ~\cite{aalr79,gor79}
\begin{equation}
\frac{\delta \sigma }{\sigma _{0}}\approx -\int_{\tau }^{\tau _{\phi }}\frac{v\lambda ^{d-1}dt}{(4\pi Dt)^{d/2}}=\left\{
\begin{array}{lcr}
-c_{3}\frac{\lambda ^{2}}{\ell ^{2}}(1-\frac{\tau }{\tau _{\phi }}), & d=3
\\
-c_{2}\frac{\lambda }{\ell }\ln (\tau _{\phi }/\tau ), & d=2 \\
-c_{1}(\sqrt{\frac{\tau _{\phi }}{\tau }}-1), & d=1.
\end{array}
\right.   \label{1.7}
\end{equation}%
Here $D=\frac{1}{d}v^{2}\tau $ is the diffusion constant, $\ell =v\tau $ is the mean-free path, $\tau$ is the mean time between successive elastic collisions, $\sigma _{0}=e^{2}n\tau /m$ is the Drude
conductivity with $n$ as the particle density, and $c_{i}$ are
constants of order unity. The upper limit of the
integral is the phase relaxation time $\tau _{\phi }$, i.e., the average
time after which phase coherence is lost due to inelastic or other
phase-shifting processes.
For weak localization processes to exist at all, the inequality $\tau _{\phi
}\gg \tau $ must hold. We note that the correction in three and two
dimensions depends on the ratio of wavelength $\lambda $ to mean-free path $\ell $, and gets smaller in the limit of weak disorder, where $\lambda /\ell
\ll 1$.
In two and one dimension the correction grows large in the
limit $\tau /\tau _{\phi }\rightarrow 0$ since one expects the phase
relaxation rate $1/\tau _{\phi }$ for a system in thermodynamic equilibrium
to go to zero for $T \rightarrow 0$.
By contrast, in some cases a plateau behavior of $1/\tau _{\phi }$ as a function of temperature has been found experimentally, which gave rise to the speculation that the zero point fluctuations may cause decoherence. However, given a unique ground state, it is difficult to understand how a particle in the system may loose its phase coherence. Several physical mechanisms that may lead to a plateau of $1/\tau _{\phi }$  have been identified. For a recent discussion of these issues see Ref.\cite{Delft}.

With $\tau /\tau _{\phi }\rightarrow 0$ for $T \rightarrow 0$  the weak
localization quantum correction will be large in any system in $d=1, 2$, no matter how weak the disorder. As we will
see, this behavior signals the fact that there are no extended states in $d=1,2$ dimensions. The characteristic length $L_{\phi }$ over which a
wave packet retains phase coherence is related to $\tau _{\phi }$ by the
diffusion coefficient $L_{\phi }=\sqrt{D\tau _{\phi }}$. In systems of
restricted dimension, e.g.,  films of thickness $a$ or wires of diameter $a$, the effective dimensionality of the system with respect to localization is
determined by the ratio $L_{\phi }/a$. Namely, for $L_{\phi }\ll a$ the
system is three-dimensional (3$d$), while for $L_{\phi }\gg a$ diffusion
over time $\tau _{\phi }$ takes place in the restricted geometry of the film
or wire, and the effective dimension is therefore $2$ or $1$.

\subsection{Strong localization and the Anderson transition}

The appearance of localized states is easily understood in the limit of very
strong disorder: localized orbitals will then exist at positions where the
random potential forms a deep well. The admixture of adjacent orbitals by
the hopping amplitudes will only cause a perturbation that does not
delocalize the particle. The reason for this is that nearby orbitals will
have sufficiently different energies so that the amount of admixture is
small. On the other hand, orbitals close in energy will in general be
spatially far apart, so that their overlap is exponentially small. Thus, we
can expect the wave functions in strongly disordered systems to be
exponentially localized. Whether the particles become delocalized when the
disorder strength is reduced, is a much more complex question. In one
dimension it can be shown rigorously that all states are localized, no
matter how weak the disorder~\cite{mot61,bor63,ber74}.

In three dimensions, the accepted view is that the particles are delocalized
for weak disorder.  In general, localized and extended states of the same
energy do not coexist, since in a typical situation any small perturbation would
lead to hybridization and thus to the delocalization of a localized state. We can therefore
assume that the localized and extended states of a given energy are
separated. For increasing disorder strength $\eta $ there will
then be a sharp transition from delocalized to localized states at a
critical disorder strength $\eta _{c}$. A qualitative criterion as to when
an Anderson transition is expected in $3d$-systems has been proposed by
Ioffe and Regel \cite{iofreg}. It states that as the mean free path $\ell $\
becomes shorter with increasing disorder, the Anderson transition occurs
when $\ell $ is of the order of the wavelength $\lambda $ of the particle
(which amounts to the condition $k_{F}\ell \sim 1$ in metals, where $k_{F}$
is the Fermi wave number). As we will see later, in $1d$ or $2d$ systems $\ell $ may be much longer than the wavelength and the particles are
nonetheless localized. In fact, the relevant mean free path here is the one
with respect to momentum transfer. A similar situation exists when we fix
the disorder strength, but vary the energy $E$. Electrons in states near the
bottom of the energy band are expected to be localized even by a weakly
disordered potential, whereas electrons in states near the band center (in $d=3$) will be delocalized, provided the disorder is not too strong. Thus
there exists a critical energy $E_{c}$ separating localized from delocalized
states, the so-called mobility edge~\cite{mott74, mott79}. The electron
mobility as a function of energy is identically zero on the localized side
(at zero temperature), and increases continuously with energy separation $\mid E-E_{c}\mid $ in the delocalized, or metallic, phase. The continuous
character of this quantum phase transition, termed \textit{Anderson
transition}, is a consequence of the scaling theory to be presented below.

Historically the continuous nature of the metal-insulator transition in
disordered solids has been a point of controversy for many years. According
to an earlier theory by Mott~\cite{mott74,mott79} the conductivity changes
discontinuously at the transition, such that a ``minimum metallic
conductivity'' exists on the metallic side of the transition. Numerical
simulations\cite{kmso} have shown beyond doubt that the transition is instead
continuous, at least in the absence of interactions.

In the much more complex situation of interacting electrons one finds for
the Hubbard model without disorder, using the Dynamical Mean-Field Theory
(DMFT), that the Mott-Hubbard metal-insulator transition is discontinuous at
finite temperatures, and that it becomes continuous in the limit $T\rightarrow 0$ \cite{georges96,kotv}. For the Hubbard model in the presence
of disorder (``Anderson-Hubbard model'') at $T=0$ the situation is similar:
the Mott-Hubbard metal-insulator transition is discontinuous for finite
disorder and becomes continuous in the limit of vanishing disorder \cite{Byczuk05,Byczuk10}.

\section{Fundamental theoretical concepts of Anderson localization}

\label{Fundamental}

The Anderson localization transition is a quantum phase transition, i.e., it
is a transition at zero temperature tuned by a control parameter, e.g., the
disorder strength, particle energy, or wave frequency. Unlike other quantum
phase transitions, the Anderson transition does not have an obvious order
parameter. Nonetheless, there exists a dynamically generated length scale,
the localization or correlation length $\xi $, which tends to infinity as
the transition is approached. Therefore, by drawing an analogy with magnetic
phase transitions, Wegner early on proposed scaling properties \cite{weg76}.
Later, he formulated a field-theoretic description of the Anderson
transition in the form of a non-linear sigma model (NL$\sigma $M) of
interacting matrices (rather than vectors, as for magnetic systems) \cite{weg79}. The NL$\sigma $M was later formulated in the mathematically more
tractable supersymmetric form\cite{efet97}.

\subsection{Scaling theory of the conductance}

Wegner~\cite{weg76} argued that the Anderson localization transition should
be described in the language of critical phenomena of continuous (quantum)
phase transitions. This requires the assumption of a correlation length $\xi
$ diverging as a function of disorder strength $\eta $ at the critical point
\begin{equation}
\xi (\eta )\sim \mid \eta -\eta _{c}\mid ^{-\nu }.  \label{2.2}
\end{equation}%
The conductivity is then expected to obey the scaling law
\begin{equation}
\sigma (\eta )\sim \xi ^{2-d}\sim (\eta _{c}-\eta )^{s};\;\;\eta <\eta
_{c},\;\;d>2.  \label{2.3}
\end{equation}%
This follows from the fact that $\sigma $, written in units of $e^{2}/(2\pi
\hbar )$, has dimension (1/length)$^{d-2}$, and the only characteristic
length near the transition is the correlation length $\xi $. By comparing
the conductivity exponent $s$ with the exponent of $\xi $ one finds
\begin{equation}
s=\nu (d-2).  \label{2.3a}
\end{equation}%
On the other hand, the conductance $g$ of a d-dimensional cube of length $L$, which for a good metal of conductivity $\sigma $ is given by $g(L)=\sigma
L^{d-2}$, must obey the scaling property
\begin{equation}
g(\eta ;L)=\Phi (L/\xi ).  \label{2.4}
\end{equation}%
This means that $g$ is a function of a single parameter $L/\xi $, so that
each value of $L/\xi $ corresponds to a value $g$.

\subsection{Renormalization group equation}

It then follows that $g(L)$ obeys the renormalization group (RG) equation
\begin{equation}
\frac{d\ln g}{d\ln L}=\beta (g),  \label{2.5}
\end{equation}%
where $\beta (g)$ is a function of $g$ only, and does not depend on
disorder. In a landmark paper, Abrahams, Anderson, Licciardello and
Ramakrishnan~\cite{aalr79} proposed the above equation and calculated the $\beta $-function in the limits of weak and strong disorder. A confirmation
of the assumption of scaling was obtained from a calculation of the
next-order term~\cite{gor79}.

At strong disorder one expects all states to be localized, with average
localization length $\xi $. It then follows that $g(L)$ is an exponentially
decreasing function of $L$:
\begin{equation}
g(L)\sim \exp (-L/\xi ).  \label{2.6}
\end{equation}%
In comparison with the ohmic dependence $g\sim L^{d-2}$ this is a very
non-ohmic behavior. The $\beta $-function is then given by
\begin{equation}
\beta (g)\sim \ln (g/g_{c})<0.  \label{2.7}
\end{equation}%
At weak disorder one finds from $g\sim L^{d-2}$ that
\begin{equation}
\beta (g)=d-2.  \label{2.8}
\end{equation}%
The important question of whether the system is delocalized (metal) or
localized (insulator) may be answered by integrating the RG equation from
some starting point $L_{0}$, where $g(L_{0})$ is known. Depending on whether
$\beta (g)$ is positive or negative along the integration path, the
conductance will scale to infinity or to zero, as $L$ goes to infinity.

In $d=3$ dimensions one has $\beta(g) > 0$ at large $g$, but $\beta(g) < 0$
at small $g$. Thus, there exists a critical point at $g=g_c$, where $\beta(g_c)=0$, separating localized and delocalized behavior.

On the other hand, in $d=1$ dimension one has $\beta(g) < 0$ at large and
small $g$, and by interpolation also for intermediate values of $g$, so that
there is no transition in this case and all states are localized.

The dimension $d=2$ apparently plays a special role, as in this case $\beta
(g)\rightarrow 0$ for $g\rightarrow \infty $. In order to determine whether $\beta >0$ or $<0$ for large $g$ one has to calculate the scale dependent
(i.e., $L$-dependent) corrections to the Drude result at large $g$. This is
precisely the weak localization correction already mentioned above. For a
system of finite length $L<L_{\phi }$ we should replace $\frac{1}{\tau
_{\phi }}=DL_{\phi }^{-2}$ in eq. \eqref{1.7} by $DL^{-2}$, leading to
\begin{equation}
g(L)=\sigma _{0}-a\;\;\ln \Big(\frac{L}{\ell }\Big),  \label{2.9}
\end{equation}%
where a diagrammatic calculation~\cite{aalr79} gives $a=2/\pi $ and $\sigma
_{0}=\ell /\lambda _{F}$ (in units of $e^{2}/\hbar $; $\lambda _{F}$ is the
Fermi wave length) has been used. It follows that
\begin{equation}
\beta (g)=-\frac{a}{g},\;\;\;d=2,  \label{2.10}
\end{equation}%
so that we can expect $\beta (g)<0$ for all $g$, implying that again all
states are localized. This result is valid for the \textquotedblleft
usual\textquotedblright\ type of disorder, i.e., in case all symmetries, in
particular time reversal symmetry (required for the weak localization
correction to be present) are preserved. If time-reversal invariance is
broken, e.g., by spin-flip scattering at magnetic impurities, the weak
localization effect is somewhat reduced in dimensions $d=2+\epsilon$, $\epsilon <<1$, but is not  completely removed. The
first correction term in the $\beta $-function is then proportional to $-1/g^{2}$ (see, e.g., Ref.\cite{efet97}) implying that all states are still
localized   (in $d=3$ dimensions the leading correction term is again $\sim 1/g$; see Ref.\cite{NMW10}). In the presence of a magnetic field the situation is more
complex, since the scaling of the Hall conductance is coupled to
the scaling of $g$. As a result, one finds exactly one extended state per
Landau energy level, which then gives rise to the quantum Hall effect~\cite{pru87}. On the other hand, if spin-rotation invariance is broken, but
time-reversal invariance is preserved, as is the case of spin-orbit
scattering, the correction term is proportional to $+1/g$, i.e., it is \emph{anti-localizing}. In this case the $\beta $-function in $d=2$ dimensions has
a zero, implying the existence of an Anderson transition~\cite{hln80}.

\subsection{Critical exponents}

In the neighborhood of the critical point at $g=g_{c}$ in $d=3$ we may
expand the $\beta $-function as
\begin{equation}
\beta (g)=\frac{1}{y}\Big[\frac{g-g_{c}}{g_{c}}\Big],\;\;\mid g-g_{c}\mid
\ll g_{c}.  \label{2.11}
\end{equation}%
Integrating the RG equation for $g>g_{c}$ from $g(\ell )=g_{0}$ to $\beta
\rightarrow 1$ at large $L$ we find $g(L)=\sigma L$, where
\begin{equation}
\sigma \sim \frac{1}{\ell }(g(\ell )-g_{c})^{y}.  \label{2.12}
\end{equation}%
Since $[g(\ell )-g_{c}] \propto (\eta _{c}-\eta) $, we conclude that the
inverse of the slope of the $\beta $-function, $y$, is equal to the
conductivity exponent $s=y$.

Similarly, one finds on the localized side $(g<g_{c})$
\begin{equation}
g(L)\sim g_{c}\exp \Big[-c(g_{c}-g(\ell ))^{y}L/\ell \Big]\sim g_{c}\exp
(-L/\xi ),  \label{2.13}
\end{equation}
from which the localization length follows as
\begin{equation}
\xi \sim \ell |\eta -\eta _{c}|^{-y}.  \label{2.14}
\end{equation}
The critical exponent $\nu $ governing the localization length is therefore $\nu =y=s$ in $d=3$ dimensions.

Since the critical conductance $g_{c}=O(1)$ in $d=3$, there exist no
analytical methods to calculate the $\beta $-function in the critical region
in a quantitative way. A perturbative expansion in $2+\epsilon $ dimensions,
where $g_{c}\gg 1$, is possible, but the expansion in $\epsilon $ is not
well-behaved, so that it cannot be used to obtain quantitative results for $s
$ and $\nu $ in $d=3$. There exist, however, reliable results on $\nu $ from
numerical studies, according to which $s=\nu =1.58\pm 0.02$ \cite{kmso,
slev99}.

\subsection{Dynamical scaling}

The dynamical conductivity $\sigma (\omega )$, i.e., the a.c. conductivity
at frequency $\omega $, in the thermodynamic limit in $d=3$ obeys the
scaling law \cite{sh81, igb82}
\begin{equation}
\sigma (\omega ;\eta )=\frac{1}{\xi }\Phi (L_{\omega }/\xi ),  \label{2.15}
\end{equation}%
where the scaling function $\Phi $ has been introduced in eq.~\eqref{2.4}.
Here $L_{\omega }$ is the typical length which a wave packet travels in the
time of one cycle, $1/\omega $. Since the motion is diffusive it obeys $L_{\omega }=\sqrt{D(\omega )/\omega }$. It is important to note that the
diffusion coefficient $D(\omega )$ is energy scale dependent and  is related
to the conductivity via the Einstein relation
\begin{equation}
\sigma (\omega )=\hbar N(E)D(\omega ),  \label{2.16}
\end{equation}%
where $N(E)$ is the density of states at the particle energy $E$.

At the Anderson transition, where $\xi \rightarrow \infty $, we expect $\sigma (\omega )$ to be finite. It follows that $\lim_{\xi \rightarrow
\infty }\Phi (L_{\omega }/\xi )\sim \xi /L_{\omega }$ and consequently
\begin{equation}
\sigma (\omega ;\eta )\sim \frac{1}{L_{\omega }}\;\;,\;\;\eta =\eta _{c}.  \label{2.17}
\end{equation}%
This is a self-consistent equation for $\sigma (\omega )$, with solution
\begin{equation}
\sigma (\omega )\sim \omega ^{1/3}\;\;,\;\;\eta =\eta _{c}\text{.}
\label{2.17a}
\end{equation}%
More precisely, in the above expressions $\omega $ should be replaced by the
imaginary frequency $-i\omega $, such that $\sigma (\omega )$ is a
complex-valued quantity.

In a more general notation, introducing the dynamical critical exponent $z$
by $\sigma (\omega )\sim \omega ^{1/z}$, we conclude that $z=3$. The
dynamical scaling is valid in a wide neighborhood of the critical point,
defined by $\omega >\frac{1}{\tau }(\xi /\ell )^{-z}\sim \mid \eta -\eta
_{c}\mid ^{\nu z}$, where $\nu z\approx 4.8$. This scaling regime is
accessible in experiment, not only by measuring the dynamical conductivity
directly, but also by observing that at finite temperature the scaling in $\omega $ is cut off by the phase relaxation rate $1/\tau _{\phi }$~\cite{igb82}. Therefore, assuming a single temperature power law $1/\tau _{\phi
}\sim T^{p}$, one finds the following scaling law for the temperature
dependent d.c. conductivity
\begin{equation}
\sigma (T;\eta )\sim T^{p/3}\Phi _{T}(\xi T^{p/3}).  \label{2.18}
\end{equation}
Using this scaling law one may in principle determine the critical exponent $\nu $ from the temperature dependence of the conductivity in the vicinity of
the critical point. In the case of disordered metals or semiconductors,
where studies of this type have been performed, the effect of
electron-electron interaction has to be taken into account. One major
modification in the above is that the Einstein relation is changed. Namely,
the single-particle density of states (which is not critical) is replaced by
the compressibility $\partial n/\partial \mu $, with $n$ as the density and $\mu $ as the chemical potential, which  in the presence of the long-range Coulomb interaction  is expected to vanish at the
transition, i.e., the system becomes incompressible. Another change is that
the frequency cutoff is given by the temperature. The critical exponents
determined from experiment vary widely, from $s=0.5$ (Ref.\cite{paa82}) and $s=1$ (Ref.\cite{fiel85}) to $s=1.6$ (Ref.\cite{bog99,waff99}), and from $z=2$
(Ref.\cite{bog99}) to $z=2.94$ (Ref.\cite{waff99}).

\section{Renormalized perturbation theory of quantum transport in disordered
media}

The field-theoretic description in terms of the nonlinear $\sigma $ model (NL$\sigma $M) mentioned in the beginning of Sec.~\ref{Fundamental} is believed
to be an exact framework within which the critical properties of the
Anderson transition may be, in principle,  calculated exactly. The mapping
of the initial microscopic model onto the NL$\sigma $M requires a number of
simplifications, so that the noncritical properties like the critical
disorder $\eta _{c}$, the behavior in anisotropic systems, or systems of
finite extension are no longer well represented by this model. In addition,
it is not known how to solve the NL$\sigma $M in cases of major interest,
such as in $d=3$ dimensions.

It is therefore useful to consider approximation schemes, which on one hand
keep the information about the specific properties of the system and on the
other hand account approximately for the critical properties at the
transition. Such a scheme, the self-consistent theory of Anderson localization, is available at least for the orthogonal ensemble
(in which both, time reversal and spin rotation symmetry are conserved).
This approach has been developed by us in Refs.\cite{voll80a,voll80} and was reviewed
in Ref.\cite{voll92}. It may be termed ``self-consistent
one-loop approximation'' in the language of renormalization
group theory but has, in fact, been derived following a somewhat different
logic as will be discussed below.

The appropriate language to formulate a microscopic theory of quantum
transport or wave transport in disordered media is a renormalized
perturbation theory in the disorder potential. The building blocks of this
theory for the model defined by eq. (2) are (i) the renormalized
one-particle retarded (advanced) Green's functions averaged over disorder
\begin{equation}
G_{\bm{k}}^{R,A}(E)=\Big[E-k^{2}/2m-\Sigma _{\bm{k}}^{R,A}(E)\Big]^{-1},
\label{3.1}
\end{equation}%
where $\Sigma _{\bm{k}}^{R}(E)=(\Sigma _{\bm{k}}^{A}(E))^{\ast }$ is the
self-energy, and (ii) the random potential correlator $\langle V^{2}\rangle $. The self-energy $\Sigma $ is a non-critical quantity and can be
approximated by $\Sigma _{\bm{k}}^{R}(E)\simeq -i/2\tau $, where $1/\tau $
is the momentum relaxation rate entering the Drude formula of the
conductivity and  isotropic scattering is assumed.

The quantity of central interest here is the diffusion coefficient $D$. It
follows from very general considerations~\cite{for75} that the
density-response function describing the change in density caused by an
external space and time dependent chemical potential
is given by
\begin{equation}
\chi (\bm{q},\omega )=\frac{D(\bm{q},\omega )q^{2}}{-i\omega +D(\bm{q},\omega )q^{2}}\chi _{0},  \label{3.2}
\end{equation}
where $D(\bm{q},\omega )$ is a generalized diffusion coefficient. The static
susceptibility (which is non-critical in the model of non-interacting
particles) is given by $\chi _{0}=N_{F}$ , where $N_{F}$ is the density of
states at the Fermi level. The form of $\chi $ is dictated by particle
number conservation and may be expressed in terms of $G^{R,A}$ as
\begin{equation}
\chi (\bm{q},\omega )=-\frac{\omega }{2\pi i}\sum_{\bm{k},\bm{k}{^{\prime }}}\Phi _{\bm{k}\bm{k}{^{\prime }}}(\bm{q},\omega )+\chi _{0}.  \label{3.3}
\end{equation}%
The two-particle quantity
\begin{equation}
\Phi _{\bm{k}\bm{k}{^{\prime }}}(\bm{q},\omega )=\left\langle G_{k_{+},k_{+}^{\prime}}^{R}G_{k_{-},k_{-}^{\prime}%
}^{A}\right\rangle,
\end{equation}
where
$G_{k,k^{\prime}}^{R,A}$  are non-averaged single-particle Green's functions,
$k_{\pm }=(\bm{k}\pm \bm{q}/2$, $E\pm \omega /2)$, and the angular brackets denote averaging over disorder,
may be written in terms of the irreducible vertex function $U$ as
\begin{equation}
\Phi _{\bm{k}\bm{k}{^{\prime }}}(\bm{q},\omega )=G_{k_{+}}^{R}G_{k_{-}}^{A}\Big[\delta _{\bm{k},\bm{k}{^{\prime }}}+\sum_{\bm{k}{^{\prime \prime }}}U_{\bm{k}\bm{k}{^{\prime \prime }}}(\bm{q},\omega
)\Phi _{\bm{k}{^{\prime \prime }}\bm{k}{^{\prime }}}(\bm{q},\omega )\Big].
\label{3.3a}
\end{equation}
In a diagrammatic
formulation the vertex function $U$ is given by the sum of all particle-hole
irreducible diagrams of the four-point vertex function. By expressing $G^{R}G^{A}$ as
\begin{equation}
G_{k_{+}}^{R}G_{k_{-}}^{A}=\frac{\Delta G_{\bm{k}}}{\omega -\bm{k}\cdot \bm{q}/m-\Delta \Sigma _{\bm{k}}},  \label{3.4}
\end{equation}%
where $\Delta G_{\bm{k}}=G_{k_{+}}^{R}-G_{k_{-}}^{A}$ and $\Delta
\Sigma _{\bm{k}}=\Sigma _{k_{+}}^{R}-\Sigma _{k_{-}}^{A}$ one may
rewrite eq.~\eqref{3.3a} in the form of a kinetic equation
\begin{equation}
\Big(\omega -\frac{\bm{k}\cdot \bm{q}}{m}-\Delta \Sigma _{\bm{k}}\Big)\Phi _{\bm{kk^{\prime }}}=-\Delta G_{\bm{k}}\Big[\delta _{\bm{k}\bm{k}{^{\prime }}}+\sum_{\bm{k}{^{\prime \prime }}}U_{\bm{k}\bm{k}{^{\prime \prime }}}\Phi _{\bm{k}{^{\prime \prime }}\bm{k}{^{\prime }}}\Big].  \label{3.5}
\end{equation}%
By summing eq.~\eqref{3.5}  over $\bm{k},\bm{k}{^{\prime }}$ one finds the
continuity equation
\begin{equation}
\omega \Phi (\bm{q},\omega )-q\Phi _{j}(\bm{q},\omega )=2\pi iN_{F}
\label{3.6}
\end{equation}
with the density-relaxation function
\begin{equation}
\Phi (\bm{q},\omega )=\sum_{\bm{k},\bm{k}{^{\prime }}}\Phi _{\bm{k}\bm{k}{^{\prime }}}(\bm{q},\omega ),  \label{3.7}
\end{equation}
and the current-density relaxation function
\begin{equation}
\Phi _{j}(\bm{q},\omega )=\sum_{\bm{k},\bm{k}{^{\prime }}}\frac{\bm{k}\cdot \bm{\hat{q}}}{m}\Phi _{\bm{k}\bm{k}{^{\prime }}}(\bm{q},\omega ),
\label{3.7a}
\end{equation}
where $\bm{\hat{q}}=\bm{q}/\mid \bm{q}\mid $.
Here the Ward identity $\Delta \Sigma _{\bm{k}}=\sum_{\bm{k}{^{\prime }}}U_{\bm{k}\bm{k}{^{\prime }}}\Delta G_{\bm{k}{^{\prime }}}=\sum_{\bm{k}{^{\prime
}}}U_{\bm{k}{^{\prime }}\bm{k}}\Delta G_{\bm{k}{^{\prime }}}$ has been used~\cite{voll80}. Since the Ward identity  plays a central role in the
derivation of the self-consistent equation, we provide a short proof which
does not rely on the perturbation expansion employed in Ref.\cite{voll80}.
Instead the proof follows the derivation of a similar Ward identity in the
case of wave propagation in disordered media \cite{bara95}. Starting from
the equations of motion of the single particle Green's function before
impurity averaging
\begin{equation}
\Big[E+\frac{\omega }{2}+i0+\frac{1}{2m}\nabla _{\bm{r}_{1}}^{2}-V(\bm{r}_{1})\Big]G^{R}(\bm{r}_{1},\bm{r}_{2};E+\frac{\omega }{2})=\delta (\bm{r}_{1}-\bm{r}_{2}),
\end{equation}

\begin{equation}
\Big[E-\frac{\omega }{2}-i0+\frac{1}{2m}\nabla _{\bm{r}_{3}}^{2}-V(\bm{r}_{3})\Big]G^{A}(\bm{r}_{3},\bm{r}_{4};E-\frac{\omega }{2})=\delta (\bm{r}_{3}-\bm{r}_{4})
\end{equation}
we multiply the first of these equations by $G^{A}(\bm{r}_{3},\bm{r}_{4};E-\frac{\omega }{2})$ and the second by $G^{R}(\bm{r}_{1},\bm{r}_{2};E+\frac{\omega }{2})$ and take the difference.
We now perform the limit $\bm{r}_{1}\rightarrow \bm{r}_{3}$ , upon which the
terms containing the disorder potential $V(\bm{r}_{i}),\, i=1,3,$ cancels out.
Finally, the disorder average is taken and the result is Fourier transformed
into momentum space, with the result
\begin{equation}
\sum_{\bm{k}}(\omega -\frac{\bm{k}\cdot \bm{\hat{q}}}{m})\Phi_{\bm{k}\bm{k}{^{\prime }}}(\bm{q},\omega )=G_{k_{+}^{\prime}}^{R}-G_{k_{-}^{\prime}}^{A}.
\end{equation}%
Comparing with eq.~\eqref{3.5} it is seen that the Ward identity indeed
holds.

In the hydrodynamic limit, i.e., $\omega \tau \ll 1$, $q\ell \ll 1$, the
current density is proportional to the gradient of the density, which is
expressed in Fourier space by
\begin{equation}
\Phi _{j}+iqD(\bm{q},\omega )\Phi =0.  \label{3.8}
\end{equation}
In fact, multiplying eq.~\eqref{3.5} by $\bm{k}\cdot \bm{\hat{q}}/m$ and
summing over $\bm{k}$ and $\bm{k}{^{\prime }}$, one may derive relation
\eqref{3.8} and by comparison finds
\begin{equation}
D_{0}/D(\bm{q},\omega )=1-\eta \frac{2E}{mn}\sum_{\bm{k},\bm{k^{\prime }}}(\bm{k}\cdot \bm{\hat{q}})G_{k_{+}}^{R}G_{k_{-}}^{A}U_{\bm{kk^{\prime }}}(\bm{q},\omega )G_{k^{\prime }_{+}}^{R}G_{k^{\prime }_{-}}^{A}(\bm{k^{\prime}}\cdot \bm{\hat{q}}),  \label{3.9}
\end{equation}%
where  $\eta =\pi N_{F}\langle V^{2}\rangle =\frac{1}{2\pi E\tau }$ is the disorder parameter, and $D_{0}=\frac{1}{d}v^{2}\tau $ is the bare diffusion constant.

As the Anderson transition is approached the left-hand-side of eq.~\eqref{3.9} will diverge for $q,\omega \rightarrow 0$, and therefore the
irreducible vertex $U$ has to diverge, too. The leading divergent
contribution to $U$ is given by the set of diagrams obtained by using the
following property of the full vertex function $\Gamma $ (the sum of all
four-point vertex diagrams) in the presence of time-reversal symmetry\cite{voll80a,voll80}:
\begin{equation}
\Gamma _{\bm{k}\bm{k}{^{\prime }}}(\bm{q},\omega )=\Gamma _{(\bm{k}-%
\bm{k}{^{\prime }}+\bm{q})/2,(\bm{k}{^{\prime }}-\bm{k}+\bm{q})/2}(\bm{k}+%
\bm{k}{^{\prime }},\omega ).  \label{3.10}
\end{equation}%
This relation follows if one twists the particle-hole (p-h) diagrams of $\Gamma $ such that the lower line has its direction reversed, i.e., the
diagram becomes a particle-particle (p-p) diagram. Now, if time-reversal
symmetry holds, one may reverse the arrow on the lower Green's function
lines if one lets $\bm{k}\rightarrow -\bm{k}$ at the same time. This
operation transforms p-p-diagrams back into p-h diagrams, so that an
identity is established relating each diagram of $\Gamma $ to its
transformed diagram $\Gamma ^{T}$, which yields the above relation.

The leading singular diagrams of $\Gamma $ give rise to the diffusion pole
\begin{equation}
\Gamma _{D}=\frac{1}{2\pi N_{F}\tau ^{2}}\;\;\frac{1}{-i\omega +Dq^{2}},
\label{3.11}
\end{equation}%
where $D$ is the renormalized diffusion coefficient. These diagrams are of
the ladder-type and therefore reducible. Their transformed counterparts $\Gamma _{D}^{T}$ are, however, irreducible and thus contribute to $U$. We
may therefore approximate the singular part of $U$ by
\begin{equation}
U_{\bm{kk^{\prime }}}^{\mathrm{sing}}=\frac{1}{2\pi N_{F}\tau ^{2}}\;\;\frac{%
1}{-i\omega +D(\bm{k}+\bm{k}{^{\prime }})^{2}}.  \label{3.12}
\end{equation}%
In low-order perturbation theory $U^{\mathrm{sing}}$ is given by the
``maximally crossed diagrams'', which when
summed up give a result $U^{\mathrm{sing},0}$ similar to eq.~\eqref{3.12},
with $D$ replaced by the diffusion constant $D_{0}$. When $U^{\mathrm{sing},0}$ is substituted as a vertex correction into the conductivity
diagram, the result is exactly the weak-localization correction discussed in
Sec.~\ref{Weak-localization}. The structure of the kernel $U_{\bm{kk^{\prime
}}}$ has been analyzed from a general viewpoint in Ref.\cite{sus07}. The
importance of the diffusion pole for the Anderson localization problem was
discussed in Ref.\cite{Janis1,Janis2} in connection with the derivation of mean-field theories for
disordered systems in the limit of high spatial dimensions.

\section{Self-consistent theory of Anderson localization}

It follows from eq.~\eqref{3.9} that for $d\leq 2$ even the lowest-order
correction in the disorder parameter $\eta $ to the inverse diffusion
coefficient (obtained by replacing $U_{\bm{kk^{\prime }}}$ by $U^{\mathrm{sing},0}$) yields a contribution which, in principle, diverges in the limit $\omega \rightarrow 0 $. This infrared divergence depends crucially on the
dimension $d$ and leads to a breakdown of perturbation theory in dimensions $d\leq 2$. In higher dimensions the divergence takes place at finite disorder
strength. Since 
the fundamental reason for the divergence of $D_{0}/D(0,0)$, eq.~\eqref{3.9}, is the presence of diffusion poles in the kernel $U_{\bm{kk^{\prime }}}$,
and
since these diffusion poles depend on the renormalized diffusion
coefficient, Vollhardt and W\"{o}lfle\cite{voll80a,voll80} interpreted eq.~\eqref{3.9} as a self-consistent equation for the diffusion coefficient.

By construction eq.~\eqref{3.9} is in agreement with perturbation theory. An
earlier attempt to set up a self-consistent equation in the spirit of
mode-mode coupling theory \cite{gpw79} failed to reproduce the weak
localization results, as it did not account for quantum interference
effects. A later \emph{ad hoc} modification of the latter theory led to a
self-consistency scheme\cite{pr81, begogo81} which is in partial agreement
with the one presented here, the main difference being that an additional
classical (i.e., not interference related) mechanism of localization is
included.

When $U^{\mathrm{sing}}$ from eq.~\eqref{3.12} is substituted for $U$, eq.~\eqref{3.9} for the diffusion coefficient $D(\omega )$ (i.e., in the limit $q\rightarrow 0$) leads to the following self-consistent equation for the
frequency-dependent diffusion coefficient $D(\omega )$:\cite{voll80a,voll80}
\begin{equation}
\frac{D_{0}}{D(\omega )}=1+\frac{k_{F}^{2-d}}{\pi m}\int_{0}^{1/\ell }\;dQ\frac{Q^{d-1}}{-i\omega +D(\omega )Q^{2}}.  \label{3.13}
\end{equation}%
Here we assumed that a finite limit $\lim_{\bm{q}\rightarrow 0}D(\bm{q},\omega )=D(\omega )$ exists, and that $Q$ is limited to $1/\ell $ in the
diffusive regime.

Eq.~\eqref{3.13} may be re-expressed as
\begin{equation}
\frac{D(\omega )}{D_{0}}=1-\eta dk_{F}^{2-d}\int_{0}^{1/\ell }\;dQ\frac{Q^{d-1}}{-i\omega /D(\omega )+Q^{2}}.  \label{3.14}
\end{equation}

\subsection{Results of the self-consistent theory of Anderson localization}

In $d=3$ eq. (3.14) has a solution in the limit $\omega
\rightarrow 0$ up to a critical disorder strength $\eta _{c}$
\begin{equation}
D=D_{0}(1-\frac{\eta }{\eta _{c}}),\;\;\eta <\eta _{c}=\frac{1}{\sqrt{3\pi }},  \label{3.15}
\end{equation}
which implies the critical exponent of the conductivity $s=1$. The $\omega $-dependence of $D(\omega )$ at the critical
point is obtained as~\cite{shap82}
\begin{equation}
D(\omega )=D_{0}(\omega \tau )^{1/3},\;\;\eta =\eta _{c},  \label{3.15a}
\end{equation}
implying a dynamical critical exponent $z=3$ in agreement with the exact result of Wegner\cite{weg76}.

At stronger disorder, $\eta >\eta _{c}$, all states are found to be
localized. The localization length $\xi ,$ defined by $\xi
^{-2}=\lim_{\omega \rightarrow 0}(-i\omega /D(\omega )),$ is found as
\begin{equation}
\xi =\frac{\sqrt{\pi }}{2}\ell \Big|1-\frac{\eta }{\eta _{c}}\Big|^{-1},
\label{3.16}
\end{equation}%
i.e., the exponent is $\nu =1$. For general $d$ in the interval $2<d<4$ one
finds Wegner scaling, $s=\nu (d-2)$. 
An extension of the self-consistent theory with respect to the momentum dependence of the renormalized diffusion coefficient near the Anderson transition has been proposed in Ref.~\cite{Garcia-Garcia08}. It leads to a modified critical exponent of the localization exponent, $\nu=1/(d-2)+1/2$, which is in much better agreement with numerical results in $d=3$. The conductivity exponent is found to be unchanged ($s=1$), i.e., Wegner scaling is no longer obeyed.

In dimensions $d \leq 2$, there is no metallic-type solution. The
localization length is found as
\begin{eqnarray}
\xi &=& \ell \Big[\exp \frac{1}{\eta} - 1\Big]^{1/2},\;\; d=2  \nonumber \\
\xi &\cong& c_1 \ell,\;\;\;\; d=1  \label{3.17}
\end{eqnarray}
where the coefficient $c_1 \approx 2.6$, while the exact result is $c_1 = 4$~\cite{ber74}.

\begin{figure}[tbp]
\begin{center}
\includegraphics[scale=0.4]{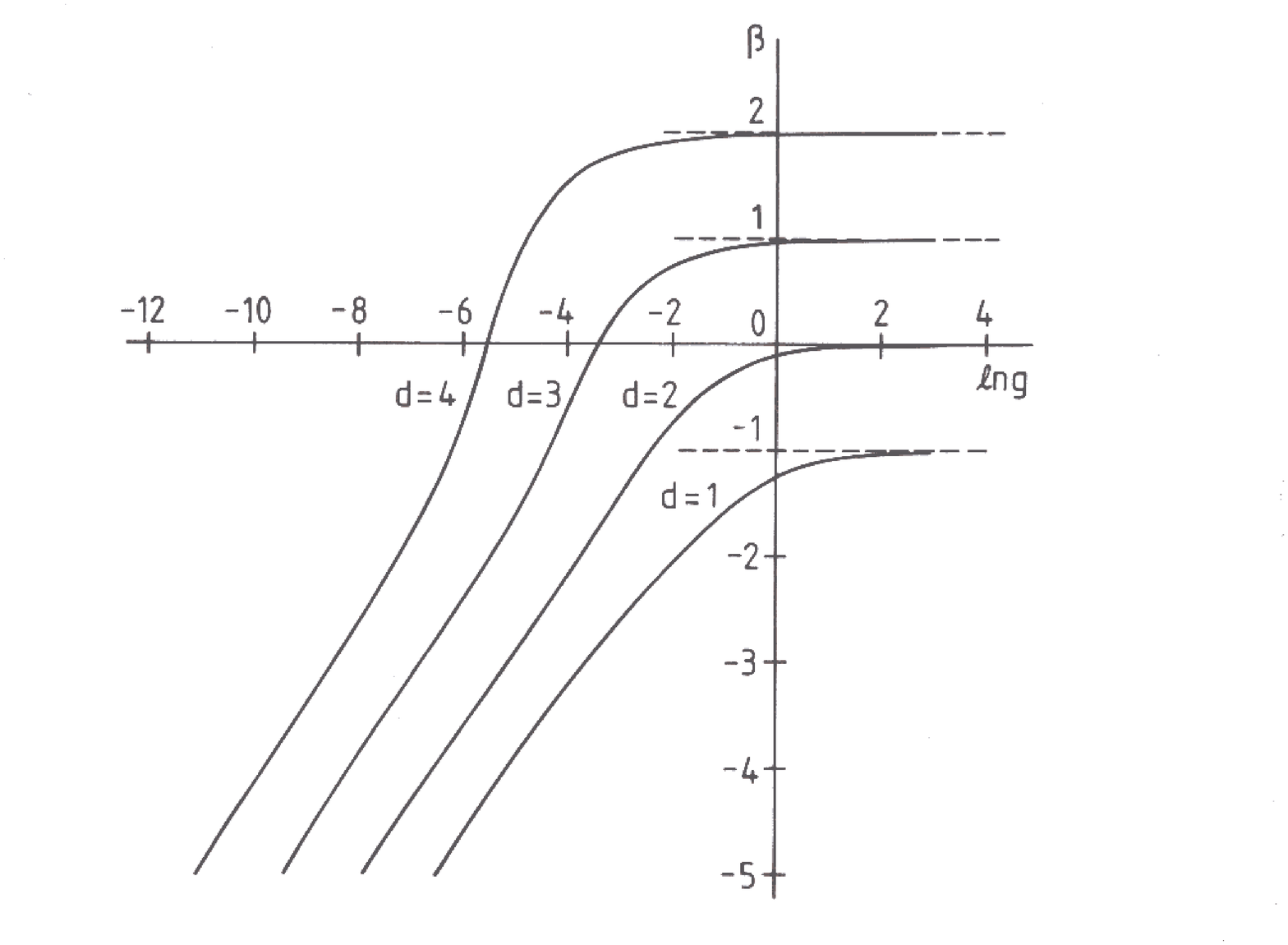}
\end{center}
\caption{Renormalization group $\protect\beta $-function in dimensions $d=1,
2, 3$ for the orthogonal ensemble, as obtained from  the self-consistent theory\cite{vollw82}.}
\label{fig1}
\end{figure}

The $\beta $-function has been derived from the self-consistent equation for
the length-dependent diffusion coefficient, where a lower cutoff $1/L$ has
been applied to the $Q$-integral in eq.~\eqref{3.14}. The result~\cite{vollw82} for $d=3$ dimensions in the metallic regime is given by
\begin{equation}
\beta (g)=\frac{g-g_{c}}{g},\;\;g>g_{c}=\frac{1}{\pi ^{2}},  \label{3.17a}
\end{equation}
and in the localized regime by
\begin{equation}
\beta (g)=1-\frac{1}{\pi ^{2}g}\frac{1+x}{1+x^{2}}e^{-x}-\frac{x^{2}}{1+x}%
\;\;,\;\;g<g_{c}.  \label{3.18}
\end{equation}
Here $x=x(g)$ is the inverse function of
\begin{equation}
g=\frac{1}{\pi ^{2}}(1+x)e^{-x}(1-x\arctan \frac{1}{x}).  \label{3.19}
\end{equation}
The $\beta $-functions in $d=1,2,3$ obtained in this way are shown in Fig.~\ref{fig1}.

\begin{figure}[tbp]
\begin{center}
\includegraphics[scale=0.4,angle=90]{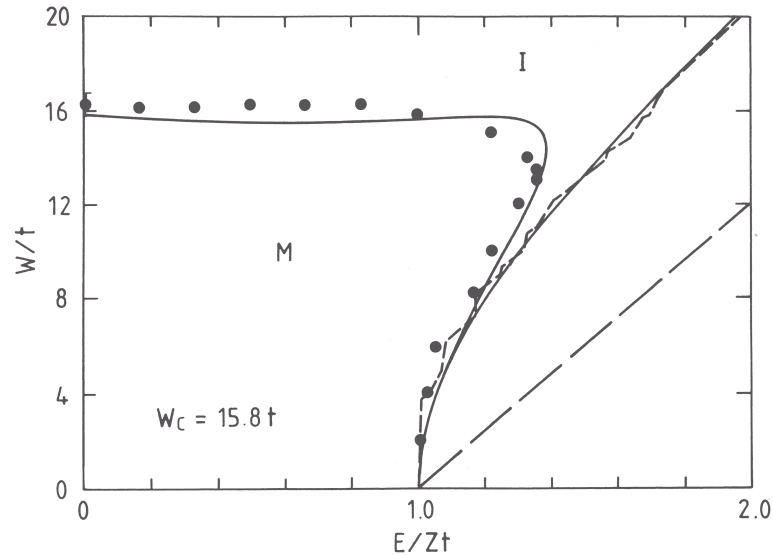}
\end{center}
\caption{Phase diagram showing metallic $(M)$ and insulating $(I)$ regions
of the tight-binding model with site-diagonal disorder (box distribution of
width $W$). Dots: numerical study~\protect\cite{kmso}; solid line:
self-consistent theory~\protect\cite{kkw90}. The remaining lines are bounds
on the energy spectrum; see Ref.\cite{voll92}. }
\label{fig2}
\end{figure}

The phase boundary separating localized and extended states in a disordered
three-dimensional system may be determined approximately by a variety of
methods. For electrons on a cubic lattice with nearest-neighbor hopping and
one orbital per site with random energy $\epsilon _{i}$ chosen from a box
distribution in the interval $[-W/2,W/2],$ the phase diagram has been
determined by numerical simulations~\cite{bul87} as shown in Fig.~\ref{fig2}.

Also shown is the result of an analytic expression obtained from the
self-consistent theory\cite{kkw90} applied to a tight-binding
model, where the Coherent Potential Approximation (CPA) was used to
evaluate the single-particle properties; no adjustable parameters enter. The
agreement is seen to be very good.

\section{Applications of the self-consistent theory of Anderson localization}

The self-consistent theory of Anderson localization proposed by us in 1980
\cite{voll80a,voll80} was applied and extended to account for many of the salient
features of disordered systems. Here we briefly review the more recent
developments, not yet described in our review\cite{voll92}. While initially
the main interest had focussed on disordered electronic systems, in recent
years the interest shifted to localization of classical waves and even more
recently, to ultracold atom systems. We first review an extension of the
self-consistent theory to the case of weak applied magnetic and electric
fields.

\subsection{Effect of static magnetic and electric fields}

\subsubsection{Magnetic fields}

One of the limitations of the self-consistent theory has been the difficulty
to treat scale dependent contributions to the conductivity in the presence
of a magnetic field in perturbation theory. As explained above, a magnetic
field induces a \textquotedblleft mass\textquotedblright\ in the Cooperon
propagator and therefore removes the localizing interference effect leading
to localization of all states in $d\leq 2$ dimensions. On the other hand
studies of the nonlinear $\sigma $-model show that in higher (two-loop)
order scale dependent terms appear which are generated solely by diffusion
propagators (diffusons). There is, however, a general theorem of
perturbation theory, related to gauge invariance, stating that the singular
contribution of any diagram with one diffuson and an arbitrary decoration
with additional impurity lines cancels within a group of related diagrams\cite{voll80}. The way out of this apparent contradiction has only been
found very recently \cite{ostr09}.

A satisfactory generalized self-consistent theory for the case of unitary
symmetry, including the two-loop and higher contributions has not been
formulated yet. Nonetheless there is a parameter regime of weak magnetic
field $B$ ($\omega _{c}=eB/mc$ ) and moderately strong disorder, $\omega
_{c}\tau <<1/\epsilon _{F}\tau \lesssim 1$ in which the one-loop
contributions still dominate over the two-loop contributions and a
generalized self-consistent theory may be formulated. The most complete
discussion of this approach was given by Bryksin and Kleinert \cite{bryksk94}, who proposed a set of two coupled self-consistent equations for the
diffusion coefficients $D_{ph}$ in the particle-hole channel (diffuson) and $D_{pp}$ in the particle-particle channel (Cooperon) of a $2d$ system:

\begin{eqnarray}
\frac{D_{pp}}{D_{0}}&=&1-g\big[\psi (\frac{1}{2}+\ell _{B}^{2}\kappa ^{2}+\frac{\ell _{B}^{2}}{4\tau _{\phi }D_{ph}})-\psi (\frac{1}{2}+\frac{\ell
_{B}^{2}}{4\tau _{\phi }D_{ph}})\big],  \label{Dpp} \\
\frac{D_{ph}}{D_{0}}&=&1-g\ln (1+\frac{\tau _{\phi }}{\tau }\frac{D_{pp}}{D_{0}}).  \label{Dph}
\end{eqnarray}

Here $\psi (z)$ is the digamma function, $g=2/(\pi k_{F}\ell )$ is the
coupling constant, $\ell _{B}=(c/eB)^{1/2}$ is the magnetic length, and $\kappa =1/(\sqrt{2}\ell )$. The solution of these equations allows one to
extend the results of weak localization theory, e.g., for the negative
magnetoresistance, to the regime of moderately strong disorder, leading to
renormalized values of the parameters of weak localization theory. Good
agreement has been found with experimental data in that range.\cite{bryksk94}

\subsubsection{Electric fields}

An applied static electric field $\bm{E}$ affects the localization physics
in the following way: electrons drifting under the influence of $\bm{E}$
experience a reduced probability of return, weakening the localization
effect provided by interference of return paths. This effect is incorporated
into the Cooperon dynamics, leading to a new term in the diffusion pole
\begin{equation}
\Gamma _{D}=\frac{1}{2\pi N_{F}\tau ^{2}}\,\,\frac{1}{-i\omega +Dq^{2}+i\mu _{d}\bm{q}\cdot\bm{E}},
\end{equation}
where $\mu _{d}=e/(m\tau )$ is the mobility. The electric field term leads
to the appearance of a localization transition even in dimension $d=1,2$ .
Near the transition in $d=1$ the diffusion coefficient is found as \cite{bryksk94}
\begin{equation}
D(E)=\left\{
\begin{array}{ll}
D_{0}(1-E_{0}/E), & \mbox{ for }E>E_{0} \\
0 & \mbox{ for }E_{0}<E,%
\end{array}
\right.   \label{D3}
\end{equation}
in agreement with the exact result in Ref.\cite{prig80}, where $E_{0}=(2n/(e\pi N_{F}^{2}D_{0})$ is the threshold field. In $d=2$ dimensions
the behavior above threshold is logarithmic:
\begin{equation}
D(E)=\left\{
\begin{array}{ll}
D_{0}\ln (E/E_{0}), & \mbox{ for }E>E_{0} \\
0 & \mbox{ for }E_{0}<E,%
\end{array}%
\right.   \label{D2}
\end{equation}%
where $E_{0}=(4\epsilon _{F}/e\kappa )\exp (-\pi k_{F}\ell /2)$. The
relaxation of the charge current following a sudden switch on of the
electric field has been considered in Ref.\cite{brykk94}. There it was found
that the current has a long time tail $\varpropto t^{-1/2}$ as a consequence
of the infrared singular behavior of the Cooperon pole.

The way in which electric and magnetic fields affect transport near the
localization transition in anisotropic systems was studied in Refs.\cite{kleinb95, brykk96}.

\subsection{Anisotropic systems, films and wires}

The question of how the scaling properties of the conductance are modified
in anisotropic systems was first addressed in Ref. \cite{wb84}. There it was
established that even in the presence of an anisotropic electronic band
structure and an anisotropic impurity-scattering cross section the
one-parameter scaling theory holds. The ratios of the components of the
conductivity tensor are invariant under scaling, implying that the geometric
mean of the conductivity components plays the role of the scaling quantity.
This feature is preserved by the self-consistent theory. Numerical studies
of anisotropic systems \cite{zsouk96, zsouk97} appeared to cast doubts on
the one-parameter scaling hypothesis. However, a later more careful study of
the problem in $d=2$ dimensions showed that indeed one-parameter scaling is
obeyed \cite{lisouk97}: the ratio of the localization lengths (in the
direction of the principal axes) turns out to be proportional to the square
root of the ratio of the conductivities. A comparison with the
self-consistent theory in the somewhat simpler form of the ``potential-well analogy'' \cite{ecsouk83} showed again
qualitative agreement. Localization in anisotropic systems has also been
considered in a model with anisotropic random potential correlations, and
the phase diagram has been mapped out within an extension of the
self-consistent theory \cite{chuz93}. The same authors explored the
consequences of finite-range correlations of the random potential within a
generalization of the self-consistent theory.\cite{chuz89}

A somewhat different but related question is the behavior of the conductance
of a film of finite thickness, or a wire of finite diameter. There is no
doubt that in the thermodynamic limit these systems behave like true $2d$ or
$1d$ systems. It is, however, interesting to understand how this behavior is
approached. Numerical studies of metallic disordered films as a function of
film thickness seemed to indicate a localization transition as a function of
thickness \cite{siku02}, in contradiction to the results of the
self-consistent theory applied to this system. A further study by the same
authors\ \cite{siku04} on systems of finite thickness in a magnetic field
explored the possibility of a delocalization transition controlled by both
the thickness and the magnetic field. The transitions obtained are
pseudo-transitions marking a crossover from strong to weak localization, as
confirmed in a later more accurate numerical study \cite{cesi06}.

\subsection{Anderson localization of classical waves}

The concept of the self-consistent theory of localization can be carried
over to the case of propagation of classical waves in disordered media. Here
we sketch the formulation following the presentation of Kroha, Soukoulis, and W\"{o}lfle\cite{ksw93}.
For scalar waves propagating in a medium of randomly positioned point
scatterers of density $n_{I}=a^{-3}$, modelled by spheres of volume $V_{s}$,
the average phase velocity is given by $c_{ph}=c_{0}[1+(V_{s}/a^{3})\Delta
\epsilon ]^{-1/2}$, where $c_{0}$ is the bare phase velocity and $\Delta
\epsilon $ characterizes the strength of the scattering (\textquotedblleft
dielectric contrast\textquotedblright ). The Green's function of the wave
equation is defined as
\begin{equation}
G_{\bm{k}}(\omega )=[G_{0}^{-1}(\omega )-\Sigma _{\bm{k}}(\omega )],
\end{equation}
where $G_{0}^{-1}(\omega )=\omega ^{2}-c_{0}^{2}k^{2}$. The self-energy $\Sigma $ may be determined within the Coherent Potential Approximation (CPA)
(see, e.g., Ref.\cite{segc89}) provided it is independent of $k$. Then the
bare diffusion constant is found as
\begin{equation}
D_{0}=2c(\omega )\frac{c_{0}}{\omega }G_{0}^{-1}(\omega )\sum_{\bm{k}}^{{}}(\bm{k}\cdot \widehat{\bm{q}})^{2}(\mathrm{Im}G_{\bm{k}}^{A})^{2}.
\end{equation}
The renormalized diffusion coefficient may be shown to satisfy the
self-consistency equation
\begin{equation}
D(\Omega )=D_{0}-2[c(\omega )\frac{c_{0}}{\omega }]^{2}\frac{\mathrm{Im}\Sigma }{(\mathrm{Im}G_{0})^{2}}\frac{D(\Omega )}{D_{0}}\sum_{\bm{k},\bm{k^{\prime }}}^{{}}(\bm{k}\cdot \widehat{\bm{q}})\frac{\mathrm{Im}G_{\bm{k}}(\mathrm{Im}G_{\bm{k^{\prime }}})^{2}}{-i\Omega +D(\Omega )(\bm{k}+\bm{k^{\prime }})^{2}}(\bm{k^{\prime }}\cdot \widehat{\bm{q}}).
\end{equation}
Here $\Omega $ is the external frequency while $\omega $ is the frequency of
the waves which enter in one-particle quantities. This equation can be
solved in the limit $\Omega \rightarrow 0$ to obtain the diffusion
coefficient in the delocalized phase and the localization length $\xi
=\lim_{\Omega \rightarrow 0}[D(\Omega )/(-i\Omega )]^{1/2}$ in the localized
phase. One finds that it is much harder to localize classical waves as compared to
electrons, and there is only a narrow region of the phase diagram (at
reasonable contrast $\Delta \epsilon $ ) where localization is found\cite{ksw93}.

A more realistic theory of the propagation of electromagnetic waves in
disordered materials with loss or gain mechanisms keeping the vector
character of the fields has been worked out by Lubatsch, Kroha, and Busch\cite{lkb05}. 
We briefly sketch the main results here. The electric field amplitude $\bm{E}_{\omega }\bm{(r})$ of an electromagnetic wave of frequency $\omega $ in a
medium with random dielectric constant $\epsilon (\bm{r};\omega )=\overline{\epsilon }(\omega )+\Delta \epsilon (\bm{r};\omega )$, obeys the wave
equation
\begin{equation}
\nabla \times (\nabla \times \bm{E}_{\omega }\bm{(r}))-\frac{\omega ^{2}}{c^{2}}\epsilon (\bm{r};\omega )\bm{E}_{\omega }\bm{(r})=\omega \bm{J}_{\omega }(\bm{r}),
\end{equation}
where $\left\langle \Delta \epsilon (\bm{r};\omega )\right\rangle =0$. In
the following the random part of the dielectric function will be modelled as
$\Delta \epsilon (\bm{r};\omega )=-(c^{2}/\omega ^{2})h(\omega )V(\bm{r})$.
The Green's functions of the wave equation, after disorder averaging, are
defined as
\begin{equation}
\bm{G}_{\bm{k}}^{R,A}(\omega )=[(\frac{\omega ^{2}}{c^{2}}\overline{\epsilon
}-k^{2})\bm{P}-\bm{\Sigma }_{\bm{k}}^{R,A}(\omega )]^{-1}.
\end{equation}
Here $\bm{G}$ and the self-energy $\bm{\Sigma }$ are $(3\times 3)$ tensors
and $\bm{P}=\bm{1}-\widehat{\bm{k}}\otimes \widehat{\bm{k}}$ is the
projector onto the transverse subspace (here and in the following the hat
symbol denotes a unit vector). Transport properties are contained in the
two-particle correlation function (a tensor of rank four)
\begin{equation}
\bm{\Phi }_{\bm{kk^{\prime }}}(\bm{q},\Omega )=\left\langle \bm{G}_{k_{+},k_{+}^{\prime }}^{R}\otimes \bm{G}_{k_{-},k_{-}^{\prime
}}^{A}\right\rangle ,
\end{equation}
where $k_{\pm }=(\bm{k}\pm \bm{q}/2,\omega \pm \Omega /2)$ , etc., which
obeys the Bethe-Salpeter equation
\begin{equation}
\bm{\Phi }_{\bm{kk^{\prime }}}(\bm{q},\Omega )=\bm{G}_{k_{+}}^{R}\otimes \bm{G}_{k_{-}}^{A}\Big[\delta _{\bm{kk^{\prime }}}+\sum_{\bm{k^{\prime
\prime }}}\bm{U}_{\bm{kk}^{\prime \prime }}(\bm{q},\Omega )\bm{\Phi }_{\bm{k^{\prime \prime }}\bm{k}^{\prime }}(\bm{q},\Omega )\Big].
\end{equation}
As in the case of electrons in a random potential considered above, the
Bethe-Salpeter equation may be converted into a kinetic equation for the
integrated intensity correlation tensor $\bm{\Phi }_{\bm{k}}(\bm{q},\Omega
)=\sum_{\bm{k}^{\prime }}\bm{\Phi }_{\bm{kk^{\prime }}}(\bm{q},\Omega )$ of
the form
\begin{equation}
\Big(\Delta \bm{G}_{\bm{k},0}^{-1}(\omega )-\Delta \bm{\Sigma }_{\bm{k}}\Big)\bm{\Phi }_{\bm{k}}=-\Delta \bm{G}_{\bm{k}}\Big[\bm{1\otimes 1}+\sum_{\bm{k^{\prime \prime }}}\bm{U}_{\bm{kk^{\prime \prime }}}\bm{\Phi }_{
\bm{k^{\prime \prime }}}\Big],
\end{equation}
where $\Delta \bm{G}_{\bm{k},0}^{-1}(\omega )=[\bm{G}_{k_{+},0}^{R}]^{-1}\bm{\otimes 1}-\bm{1\otimes }[\bm{G}
_{k_{-},0}^{A}]^{-1}$, $\Delta \bm{\Sigma }_{\bm{k}}=\bm{\Sigma }_{k_{+}}^{R}\bm{\otimes 1}-\bm{1\otimes \Sigma }_{k_{-}}^{A}$, and $\Delta \bm{G}_{\bm{k}}=\bm{G}_{k_{+}}^{R}\bm{\otimes 1}-\bm{1\otimes G}_{k_{-}}^{A}.$

The kinetic equation serves to derive the energy conservation equation and
the equivalent of Fick's law:
\begin{equation}
\lbrack \Omega +\frac{i}{\tau _{L}(\Omega )}]P_{E}(\bm{q},\Omega )+\bm{q\cdot J}_{E}(\bm{q},\Omega )=S(\bm{q},\Omega )
\label{Fick1}
\end{equation}
\begin{equation}
\bm{J}_{E}(\bm{q},\Omega )=iP_{E}(\bm{q},\Omega )\bm{D}(\Omega )\cdot \bm{q}.
\label{Fick2}
\end{equation}
Here
\begin{equation}
P_{E}(\bm{q},\Omega )=(\omega /c_{p})^{2}\sum_{\bm{k}}\bm{\Phi }_{\bm{k}}(\bm{q},\Omega )
\end{equation}
 is the energy-density relaxation function, with $c_{p}$ as the renormalized phase velocity, and
\begin{equation}
 \bm{J}_{E}(\bm{q},\Omega
)=(\omega /c_{p})\bm{v}_{E}(\omega )\sum_{\bm{k}}(\bm{k\cdot }\widehat{\bm{q}})\bm{\Phi }_{\bm{k}}(\bm{q},\Omega )
\end{equation}
is the energy-current density
relaxation function, with $\bm{v}_{E}(\omega )$ as the energy transport
velocity; for the definitions of \ $c_{p}$ and $\bm{v}_{E}(\omega )$ we
refer the reader to Ref.\cite{lkb05}. When energy absorption by the medium
is taken into account (as expressed by the imaginary part of the dielectric
function), or conversely, if  a medium with gain is considered, energy is not
conserved, as expressed by the loss/gain rate $\frac{1}{\tau _{L}(\Omega )}$. The energy-diffusion coefficient tensor $\bm{D}(\Omega )$ is found as
\begin{equation}
\bm{D}(\Omega )=\frac{1}{3}v_{E}(\omega )\bm{l}_{T},
\end{equation}
where the tensor of transport mean free path is given by
\begin{equation}
\bm{l}_{T}=\frac{c_{p}}{\omega }(A+\kappa)\bm{l}.
\end{equation}
Here the main contribution to $\bm{l}$ has a form which is analogous to eq.~\eqref{3.9}:
\begin{equation}
\bm{l}^{-1}=a_{1}^{-1}\sum_{\bm{k,k^{\prime }}}(\bm{k\cdot } \widehat{\bm{q}})\Delta \bm{G}_{\bm{k}}\bm{U}_{\bm{kk^{\prime \prime }}}(0,\Omega )\Delta \bm{G}_{\bm{k^{\prime }}}(\bm{k}^{\prime } \bm{\cdot }\widehat{\bm{q}}),
\end{equation}
and $a_{1}$ and $A$ are defined in Ref.\cite{lkb05}. The quantity $\kappa$ describes scattering caused by a mismatch of absorption/gain
between the scattering objects and the medium.

The energy density propagator $P_{E}(\bm{q},\Omega )$ in the limit of small $\bm{q},\Omega $ follows from eqs. \eqref{Fick1}, \eqref{Fick2} as
\begin{equation}
P_{E}(\bm{q},\Omega )=[\Omega +\frac{i}{\tau _{L}(\Omega )}+i\bm{q} \cdot \bm{D}
(\Omega ) \cdot \bm{q}]^{-1}S(\bm{q},\Omega).
\end{equation}
Replacing $\bm{U}_{\bm{kk}^{\prime \prime}}(0,\Omega )$ by its singular part
proportional to the diffusion propagator $P_{E}(\bm{q},\Omega )$ one arrives
at a self-consistent equation for the diffusion coefficient tensor. The
latter provides a framework for the description of  the interplay between localization and
stimulated emission in materials with gain, i.e., the problem of the random
laser\cite{flk06}.

The predictions of the self-consistent theory have also been probed by
comparison with numerical results\cite{lobw05} for transmission of waves in
unbounded $1d$ and $2d$ systems and through strips of finite width. Good
overall agreement is found.

The localization of phonons and the ultrasound attenuation in layered
crystals with random impurities has been studied within the self-consistent theory in Ref.\cite{chulz99}.

\subsection{Transport through open interfaces}

Most of the discussion so far considered transport in infinitely extended
systems, with the exception of the scaling theory for systems of length $L$.
In some cases, however, transport through plate-shaped systems in the
direction perpendicular to the plate surface is of interest. As pointed out
by van Tiggelen and collaborators \cite{vTlw00, skipvt}, the weak
localization physics changes near an open boundary, as the finite
probability of escape through the interface diminishes the return
probability necessary for interference. In the framework of the
self-consistent theory this effect may be taken into account quantitatively.
To this end it is useful to express eq.~\eqref{3.13} in position-energy
space as

\begin{equation}
\frac{D_{0}}{D(\omega )}=1+2\pi \frac{k_{F}^{2-d}}{m}C(\bm{r},\bm{r'}),
\end{equation}
where $C(\bm{r},\bm{r'})$ is a solution of the diffusion equation, and a
cut-off $Q<1/\ell$ to the momentum  was applied to the spectrum of the $Q$-modes in eq.~\eqref{3.13}
\begin{equation}
\lbrack -i\omega +D(\omega )\nabla ^{2}]C(\bm{r},\bm{r'})=\delta (\bm{r}-\bm{r'}).
\end{equation}
The above formulation now allows one to describe position-dependent
diffusion processes, as they appear near the sample surface in a confined
geometry, e.g., transmission through a slab. In that case the diffusion
coefficient may be assumed to be position dependent, $D=D(\bm{r},\omega )$.
Then $C(\bm{r},\bm{r\prime})$ obeys the modified diffusion equation~\cite{hu08}

\begin{equation}
\lbrack -i\omega +\nabla D(\bm{r},\omega )\nabla ]C(\bm{r},\bm{r'})=\delta (\bm{r}-\bm{r'}).
\end{equation}%
The solution is subject to an appropriate boundary condition at the surface
of the sample. A microscopic derivation of the above equation in
diagrammatic language was given in Ref.\cite{chersk08}. Further confirmation
of the theory was obtained in \cite{tian08}, where the above equations was
derived within the nonlinear $\sigma $-model framework in the weak coupling
limit. The theory accounts very well for the localization properties of
accoustic waves transmitted through a strongly scattering plate~\cite{hu08}.

It is natural to ask whether a position dependent diffusion coefficient will
change the critical behavior obtained from the scaling properties of the
conductance of finite size samples. This question was addressed in Ref.\cite{cher09} with the result that the critical exponents are unchanged and the $\beta $-function is hardly modified by the improved approximation. The
scaling of the transmission coefficient for classical waves through a
disordered madium near the Anderson transition was considered within the
position dependent self-consistent theory in Ref.\cite{cherskit09}

The transmission of microwave pulses through quasi-one-dimensional samples
has been measured recently and was analyzed in terms of the self-consistent
theory\cite{zhacha09}. It was found that while the self-consistent theory
can account very well for the propagation at intermediate times, it fails at
longer times when the transport occurs by hopping between localized regions.

Anderson localization of atoms in a Bose-Einstein condensate released from a
trap and subject to a random potential has been considered in the framework
of the self-consistent theory in Ref.\cite{skipsha08}. The authors show that
the scaling properties govern the dynamical behavior of the expanding atom
cloud, so that the critical exponents determine the power law in time obeyed
by the expanding cloud size.

\section{Conclusion}

Anderson localization in disordered systems continues to be a very lively
field of research. Current investigations do not concentrate so much on disordered
electrons but on  classical waves (light, electromagnetic microwaves,
acoustic waves), or ultracold atoms in the presence of disorder. Although
the fundamental concepts of Anderson localization are well understood by now,
there still remain a number of open questions. Some of them are related to
the analytical theory of critical properties near the Anderson transition.
Others concern the quantitative description of realistic materials, e.g., the
question under which conditions light or acoustic waves become
localized. The self-consistent theory of Anderson localization has
been, and will continue to be, a versatile tool  for the investigation of these problems. It allows one to incorporate the
detailed characteristics of the system such as the energy dispersion relation,
the particular form of disorder, the shape of the sample, and  loss or gain mechanisms in an efficient way. The self-consistent theory is not only
applicable to stationary transport problems, but also to dynamical situations
such as pulse propagation or the behavior after a sudden switch-on.

As Anderson localization is a wave-interference phenomenon, the limitations
of phase coherence are an important subject of study in this context. By now
Anderson localization has been observed in many different systems beyond
doubt. On the other hand, the observation of the Anderson transition itself is a much more challenging task. Here the recent investigations of classical waves and atomic matter waves offer fascinating, new
perspectives which will undoubtedly lead to a deeper understanding of the
localization phenomenon.

\section*{Acknowledgements}

We thank Vaclav Jani\v{s}, Hans Kroha, Khandker Muttalib, and Costas
Soukoulis for many fruitful discussions. Financial support by the TTR 80 of
the Deutsche Forschungsgemeinschaft is grateful acknowledged.

\printindex                         

\end{document}